\documentclass[aps,pra,twocolumn,superscriptaddress,showpacs]{revtex4}
\usepackage{dcolumn}                 
\newcolumntype{w}[1]{D{.}{.}{#1}}
\newcolumntype{.}{D{x}{}{-1}}
\newcommand*{\centt}[1]{\multicolumn{1}{c}{#1}}
\newcommand*{\cent}[1]{\multicolumn{1}{c}{$#1$}}
\usepackage{times}
\usepackage{nicefrac}
\usepackage{amsmath}
\usepackage{amsfonts}
\usepackage{amssymb}
\usepackage{amsthm}

\newcommand{\bsigma}{\vec{\sigma}}

\newcommand{\bfp}{\vec{p}}

\begin{document}
\preprint{Version 0.2}

\title{Quantum electrodynamics $m \alpha^6$ and  $m \alpha^7 \ln \alpha$ corrections 
       to the fine splitting in Li and Be$^+$}

\author{Mariusz Puchalski}
\affiliation{Faculty of Physics, University of Warsaw, Pasteura 5, 02-093 Warsaw, Poland}
\affiliation{Faculty of Chemistry, Adam Mickiewicz University, Umultowska 89b, 61-614 Pozna{\'n}, Poland}

\author{Krzysztof Pachucki}
\affiliation{Faculty of Physics, University of Warsaw, Pasteura 5, 02-093 Warsaw, Poland}

\begin{abstract}
We derive quantum electrodynamics corrections to the fine structure
in three-electron atomic systems at $m \alpha^6$ and  $m \alpha^7 \ln \alpha$ orders
and present their numerical evaluations for the Li atom and Be$^+$ ion. 

\end{abstract}

\maketitle

\section{Introduction}
The fine splitting is a difference between energies of P$_{3/2}$ and P$_{1/2}$ states.
For hydrogenic systems it can be obtained from the Dirac equation,
while for many electron systems one needs quantum electrodynamic (QED) 
theory to consistently describe correlations with relativistic effects. 
The most common many-electron Dirac-like methods 
\cite{safronovaBeplus,dereviankoCC,pergerMCDF,yerokhinLilike} are able
to achieve two significant digits at most, while experimental precision 
is about 6 significant digits \cite{brown, noerters}. 
A much more accurate description of light few-electron systems 
relies on nonrelativistic version of QED, called NRQED theory.
Relativistic, retardation, electron self-interaction, and vacuum polarization effects
can all be accounted for perturbatively by expansion of energy levels
in powers of the fine structure constant $\alpha$,
\begin{equation}
E(\alpha) = m\,\alpha^2\,{\cal E}^{(2)} + m\,\alpha^4\,{\cal E}^{(4)} + 
m\,\alpha^5\,{\cal E}^{(5)} + m\,\alpha^6\,{\cal E}^{(6)} + \ldots \label{01}
\end{equation} 
where expansion coefficients ${\cal E}^{(i)}$ may include powers of $\ln\alpha$. 
Since these  expansion coefficients are expressed 
in terms of the first- and second-order matrix elements of some operators
with the nonrelativistic wave function, the accuracy of the numerical calculation
strongly depends on the quality of this function. For example, MCHF calculations
\cite{blundellMCHF,fischerTBL,godefroidMCHF} are accurate only to three digits
because the wave function is a combination of Slater determinants and does not satisfy the cusp condition.
A much more accurate nonrelativistic wave function can be obtained by using an explicitly
correlated basis such as Hylleraas functions \cite{lit_fine_yd,yan_rel,wangHeminus,lit_fine}.
However, three-electron integrals with explicitly correlated functions are much more complicated 
than two-electron ones. Moreover, the required number of basis functions 
has to be much larger in order to achieve similar accuracy as for two-electron systems.  
So the extension of QED calculations to a three-electron system is not a simple task.
In our recent works \cite{lit_fs, noerters} we performed complete calculations
of higher-order $m \alpha^6$ and $m\,\alpha^7\,\ln\alpha$ corrections to Li and Be$^+$ $2P_{3/2} - 2P_{1/2}$ 
fine splitting.  Here  we aim to present in more detail the computational methods.

The fine structure splitting at the leading order $E^{(4)}_{\rm fs}$ is given  by the expectation value
\begin{equation}
E^{(4)}_{\rm fs} = \langle H_{\rm fs}^{(4)} \rangle \label{02}
\end{equation}
of spin-dependent operators from the Breit-Pauli Hamiltonian~\cite{bs},
\begin{eqnarray}
H_{\rm fs}^{(4)} &=& \sum_a \frac{Z\,\alpha}{4\,m^2\,r_a^3}\,\vec \sigma_a\,
\bigl[(g-1)\,\vec r_a\times\vec p_a \bigr]
 \label{03}\\ &&+
\sum_{a\neq b}\frac{\alpha}{4\,m^2\,r_{ab}^3}\,
\vec \sigma_a\bigl[g\,\vec r_{ab}\times\vec p_b
-(g-1)\,\vec r_{ab}\times\vec p_a\bigr]\nonumber\,,
\end{eqnarray}
where $g$ is the exact electron g-factor. The mean value in Eq.~(\ref{02}), 
$\langle \ldots \rangle \equiv \langle \Phi | \ldots | \Phi\rangle$ is calculated 
using the wave function $\Phi$ from the stationary Schr\"odinger equation
\begin{equation}
(H-E)\Phi = 0\label{04}
\end{equation}
with the nonrelativistic Hamiltonian $H$ in the infinite nuclear mass limit  
\begin{eqnarray}
 H &=& \sum_a \frac{\vec p_a^{\,2}}{2\,m} + V \\
 V &\equiv& \sum_a -\frac{Z\,\alpha}{r_a} + \sum_{a>b}\frac{\alpha}{r_{ab}}\label{06}
\end{eqnarray}
The Li and Be$^+$ fine structure in the leading order, including finite nuclear mass corrections, 
has  been calculated by using the Hylleraas functions in 
Refs. \cite{lit_fine_yd,lit_fine}. 
rThe high accuracy is achieved by the use of a relatively large number (about 14 000) of
these functions. All matrix elements are expressed in terms of standard and extended
Hylleraas integrals,  which are obtained with the help of recursion relations \cite{rec2,rec3}.

The situation is different with matrix elements of $m\,\alpha^6$ and higher-order
operators in the Hylleraas basis, where additional classes of complicated integrals 
appear, for which no efficient numerical algorithms are known. 
Other difficulties arise in the evaluation of the second-order matrix element
with nearly singular operators. The Green function, or equivalently the sum
over pseudo-states, requires large values of nonlinear parameters.
This causes severe problems with the numerical stability of recursive algorithms
with Hylleraas integrals. We overcome this problem by the application
of another basis set, which consists of the explicitly correlated Gaussian functions.
We have found \cite{lit_hfs,bei_hfs} that the second-order matrix elements
can be calculated with high precision when nonlinear parameters are globally optimized
and a large number of Gaussian functions is employed.

\section{Higher-order fine structure}

The $m \alpha^6$ correction $E_{\rm fs}^{(6)}$ 
to  the fine structure can be expressed 
as the sum of the first- and second-order matrix elements
with the nonrelativistic wave function,
\begin{equation}
E_{\rm fs}^{(6)} = \bigg \langle H^{(4)}\,\frac{1}{(E-H)'}\,H^{(4)} \bigg \rangle + 
          \langle H^{(6)}_{\rm fs} \rangle,\label{07}
\end{equation}
where the Breit-Pauli Hamiltonian $H^{(4)}$ is of the form \cite{bs}
\begin{eqnarray}
H^{(4)} &=& H^{(4)}_A + H^{(4)}_B + H^{(4)}_C \\ 
H^{(4)}_A &=&\sum_a \biggl\{-\frac{\vec p^{\,4}_a}{8} + \frac{ \pi\,Z}{2}\,\delta^3(r_a) \biggr\} \label{09}\\
&& +\sum_{a<b} \biggl\{\pi \, \delta^3(r_{ab}) - \frac{1}{2}\, p_a^i\,
\biggl(\frac{\delta^{ij}}{r_{ab}}+\frac{r^i_{ab}\,r^j_{ab}}{r^3_{ab}}
\biggr)\, p_b^j \biggr\}\,. \nonumber\\
H^{(4)}_B &=&\sum_a \frac{Z}{4\,r_a^3}\,
\vec\sigma_a\cdot\vec r_a\times \vec p_a \nonumber \\
&& +\sum_{a \neq b} \frac{1}{4\,r_{ab}^3} \vec \sigma_a \,\big( 2\, \vec r_{ab}\times\vec p_b 
- \vec r_{ab}\times\vec p_a \big)\,.\label{10} \\
H^{(4)}_C &=& \sum_{a<b} \frac{\sigma_a^i\,\sigma_b^j}
{4\,r_{ab}^3}\,\biggl(\delta^{ij}-3\,\frac{r_{ab}^i\,r_{ab}^j}{r_{ab}^2}\biggr)\,.\label{11}
\end{eqnarray}
The potentially singular second-order quadratic term with $H^{(4)}_A$ in Eq.~\eqref{07} 
does not contribute to fine splitting, and thus is excluded from further consideration.
$H^{(6)}_{\rm fs}$ is an effective Hamiltonian of order $m\,\alpha^6$.
Following the derivation in Refs. \cite{nrqed} and \cite{lit_fine}, $H^{(6)}_{\rm fs}$
can be represented in the following form
\begin{equation}
H^{(6)}_{\rm fs} =\sum_{i=1,7} \delta H_i\,,\label{12}
\end{equation}
\begin{eqnarray}
\delta H_{1} &=& \sum_a
\frac{3}{16\,m^4}\, p_a^2\,e\,\vec{\cal E}_a\times\vec p_a \cdot \vec\sigma_a \label{13}\\
\delta H_{2} &=& \sum_{a\neq b}
-\frac{i\,\pi}{8\,m^4}\,\vec\sigma_a\cdot\vec p_a\times\delta^3(r_{ab})\,\vec p_a  \label{14} \\
\delta H_4 &=&
\sum_a\,\frac{e}{4\,m^3}\,
\Bigl[2\,p_a^2\,\vec p_a\cdot\vec {\cal A}_{a} 
+ p_a^2\,\vec\sigma_a\cdot\nabla_a\times\vec {\cal A}_{a}\Bigr] \label{15}\\
\delta H_5 &=& \sum_a \frac{e^2}{2\,m^2}\,\vec\sigma_a \cdot \vec{\cal E}_a\times \vec {\cal A}_{a} \\
&& +\frac{i\,e}{16\,m^3}\,\Bigl[\vec {\cal A}_{a}\times\vec p_a\cdot\vec\sigma_a -
\vec \sigma_a\cdot\vec p_a\times \vec {\cal A}_{a}\,, p_a^2\Bigr]  \nonumber \\
\delta H_6 &=& \sum_a\frac{e^2}{2\,m^2}\,\vec {\cal A}_a^2 \\
\delta H_7 &=& \sum_{a\neq b}\frac{\alpha}{4\,m^2}\biggl\{
-i\,\biggl[\vec\sigma_a\times\frac{\vec
    r_{ab}}{r_{ab}},\frac{p_a^2}{2\,m}\biggr]\,
    e\,\vec{\cal E}_b \nonumber \\
&& + \biggl[\frac{p_b^2}{2\,m},
\biggl[\vec\sigma_a\times\frac{\vec
    r_{ab}}{r_{ab}},\frac{p_a^2}{2\,m}\biggr]\biggr]\,\vec p_b \biggr\} \label{18}
\end{eqnarray}
where ${\cal E}_a$ is the static electric field at the position of particle $a$
\begin{equation}
e\,\vec{\cal E}_a \equiv -\nabla_a V =  
-Z\,\alpha\,\frac{\vec r_a}{r_a^3} +\sum_{b\neq a}\alpha\,\frac{\vec r_{ab}}{r_{ab}^3}
\label{19}
\end{equation}
and ${\cal A}^i_{a}$ is the vector potential at the position of particle  $a$, 
which is produced by all other particles
\begin{equation}
e\,{\cal A}^i_{a} \equiv \sum_{b\neq a} \frac{\alpha}{2\,r_{ab}}
\biggl(\delta^{ij}+\frac{r_{ab}^i\,r_{ab}^j}{r_{ab}^2}\biggr)\,
\frac{p_b^j}{m} + \frac{\alpha}{2\,m}\frac{\bigl(\vec\sigma_b\times\vec
  r_{ab}\bigr)^i}{r_{ab}^3}\,,\label{20}
\end{equation}

In order to further improve theoretical predictions, the higher-order $m\,\alpha^7$
contribution is not neglected but instead is approximated by the numerically dominating logarithmic part. 
It is obtained from the analogous correction to the helium fine structure \cite{fsqed,yerkrp}
by dropping the $\sigma^i\,\sigma^j$ terms because they do not contribute
for states with the total electron spin $S=1/2$,
\begin{equation} \label{21}
E^{(7)}_{\rm fs, log} =   \langle H^{(7)}_{\rm fs, log}\rangle 
+2\,\Bigl\langle H^{(4)}_{B} \frac1{(E_0-H_0)'}\,H^{(5)}_{\rm log}\Bigr\rangle
\end{equation}
\begin{equation}
H^{(5)}_{\rm log} = \alpha^2\ln[(Z\,\alpha)^{-2}]\biggl[\frac{4Z}{3}\,\sum_a\delta^3(r_a) 
- \frac{7}{3}\,\sum_{b<a}\delta^3(r_{ab})\biggr]
\end{equation}
\begin{eqnarray}
H^{(7)}_{\rm fs, log} &=&
\alpha^2\ln[(Z\,\alpha)^{-2}] \,\left[
\frac{Z}{3}\,\sum_a
i\,\bfp_a\times\delta^3(r_a)\,\bfp_a\cdot\bsigma_a
  \right. \nonumber \\&& 
-\frac{3}{4}\, \sum_{b\neq a}
i\,\bfp_a\times\delta^3(r_{ab})\,\bfp_a\cdot\bsigma_a 
 \biggr] \,.\label{23}
\end{eqnarray}
The neglected higher-order corrections are the nonlogarithmic $m\,\alpha^7$ term
and the finite nuclear mass corrections to the $m\,\alpha^6$ contribution.
They will limit the accuracy of our theoretical predictions for Li and Be$^+$ fine structure.

\section{Transformation of matrix elements}

The expectation value of $H_{\rm fs}^{(6)}$ in Eq.~\eqref{12} is transformed initially
to a form convenient for numerical calculations with $^2 P$-states
\begin{eqnarray}
\delta H_{1} &=& = \frac{3}{16}\,(-Z\,Q_{1} + Q_{2}) \label{24}\\
\delta H_{2} &=& -\frac{\pi}{8}\,D_2 \label{25} \\
\delta H_4 &=& -\frac{1}{4}\,(Q_{3} + Q_{4}) \label{26}\\
\delta H_5 &=& \frac{1}{4}\,\bigl[-Z (Q_{5} + Q_{6} ) + Q_{7} + Q_{8}\bigr] + \frac{1}{8}\,\bigl(-Z P_{1} + P_{2}\bigr)
\nonumber \\ && + \frac{1}{8}\,\bigl(Q_{9} - Q_{10} - Q_{3}\bigr) - \frac{1}{16}\,P_{3}  \label{27}\\
\delta H_6 &=& -\frac{1}{4}\,\bigl(Q_{11} + Q_{12}\bigr) + \frac{1}{16}\,P_{4} \label{28}\\
\delta H_7 &=&  \frac{Z}{4}\, \bigl(  Q_{14} - Q_{15}\bigr) 
       - \frac{1}{4}\, \bigl(  Q_{17} + Q_{18}\bigr)  \label{29}\\ &&
       +  \frac{1}{4}\,\bigl(-Q_{4} + Q_{19} - Q_{20}\bigr)\nonumber
\end{eqnarray}
where $Q_i$ and $P_i$ are defined in Table \ref{tabel1}. Additionally, operators $Q_1$, $Q_2$, and $Q_4$
are transformed into the sum of the singular D-term with the Dirac-$\delta$ operator and the regular R-part.
Matrix elements with D-terms are calculated with Hylleraas, while Gaussian functions are used for R-terms, 
which ensures high numerical precision.  
\begin{table*}[htb]
\renewcommand{\arraystretch}{0.95}
\caption{Expectation values of operators for Li and Be$^+$ $2^2$P$_{J}$  states, $\langle Q \rangle = K_J \, V$ with
the additional prefactor $K_J =\{1,-1/2\}$ for $J=1/2, 3/2$, correspondingly. All digits are significant.}
\label{tabel1}
\begin{tabular}{rl@{\hspace{0.5cm}}w{5.8}w{5.8}}
\hline
\hline
\multicolumn{2}{c}{Operator} & \cent{V_{\rm Li}} & \cent{V_{\rm Be^+}} \\
\hline
$Q_{1}= $ &$  \sum_a \vec\sigma_a\,p_a^2\,\frac{\vec r_{a}}{r_{a}^3}\times\vec p_a =  - 2\,\pi\, D_{1} - R_{1} $ 
 & -0.695\,207  & -14.464\,31\\
$D_{1}= $ &$  \sum_a i \,\vec\sigma_a\, \vec p_a \times \delta^3(r_a) \vec p_a $   
 & 0.097\,730   &  2.010\,13  \\  
$R_{1}= $ &$  \sum_a i \,\vec\sigma_a\, p_a^k \,\vec p_a \times \frac{1}{r_a} \vec p_a \,p_a^k$ 	 
 & 0.082\,895   &  1.834\,29 \\
$Q_{2}= $ &$  \sum_{a,b\neq a}\vec\sigma_a\,p_a^2\,\frac{\vec r_{ab}}{r_{ab}^3}\times\vec p_a = - 2\,\pi \, D_{2} - R_{2}$   
 &-0.502\,754   & -11.065\,87  \\
$D_{2}= $ &$  \sum_{a,b \neq a} i \,\vec\sigma_a\,\vec p_a \times \delta^3(r_{ab}) \vec p_a$ 
 & 0.044\,668   &  0.980\,97 \\      
$R_{2}= $ &$  \sum_{a,b\neq a} i \,\vec\sigma_a\, p_a^k \,\vec p_a \times \frac{1}{r_{ab}} \vec p_a\,p_a^k $
 & 0.222\,098   &  4.902\,28 \\
$Q_{3}= $ &$  \sum_{a,b\neq a}\vec\sigma_a\,p_a^2\,\frac{\vec r_{ab}}{r_{ab}^3}\times\vec p_b = 2\, \pi \, D_{3} + R_{3}$ 
 & 0.000\,421   &  0.737\,15 \\
$D_{3}= $ &$  \sum_{a,b \neq a} i \, \vec\sigma_a\,\vec p_b \times \delta^3(r_{ab}) \vec p_b$   
 & 0.017\,545   &  0.369\,31 \\        
$R_{3}= $ &$  \sum_{a,b \neq a} i \, \vec\sigma_a\,p_b^k \, \vec p_b \times \frac{1}{r_{ab}} \vec p_b \,p_b^k $
 &-0.109\,834   & -1.583\,29  \\
$Q_{4}= $ &$  \sum_{a,b\neq a}\vec\sigma_a\,p_b^2\,\frac{\vec r_{ab}}{r_{ab}^3}\times\vec p_b  = 2\, \pi \, D_{3} + R_{4}$
 & 0.281\,276   &  5.677\,45 \\
$R_{4}= $ &$  \sum_{a,b \neq a} i \, \vec\sigma_a\,p_b^k \, \vec p_b \times \frac{1}{r_{ab}} \vec p_b \,p_b^k $
 & 0.171\,036   &  3.357\,01 \\ 
$Q_{5}= $ &$  \sum_{a,b\neq a}\vec\sigma_a\,\frac{1}{r_{ab}} \,\frac{\vec r_{a}}{r_{a}^3} \times \vec p_b$ 
 & 0.161\,022   &  2.122\,84 \\
$Q_{6}= $ & $  \sum_{a,b\neq a}\vec\sigma_a\,\frac{\vec r_{a} \times \vec r_{ab}}{r_{a}^3 \, r_{ab}^3}\,(\vec r_{ab} \cdot \vec p_b) $
 & 0.068\,423   &  0.858\,67 \\
$Q_{7}= $ &$  \sum_{a,b\neq a, c \neq a}\vec\sigma_a\,\frac{1}{r_{ac}} \,\frac{\vec r_{ab}}{r_{ab}^3} \times \vec p_c $
 & 0.189\,027   &  2.559\,31 \\
$Q_{8}= $ &$  \sum_{a,b\neq a, c \neq a}\vec\sigma_a\,\frac{\vec r_{ab} \times \vec r_{ac}}{r_{ab}^3 \, r_{ac}^3}\,(\vec r_{ac} \cdot \vec p_c) $
 & 0.052\,774   &  0.675\,94 \\
$P_{1}=$  &$  \sum_{a,b\neq a}  (\vec \sigma_a \times \vec \sigma_b)\,\frac{\vec r_a \times \vec r_{ab}}{r_a^3\,r_{ab}^3}  $
 & -0.066\,977  & -0.904\,13 \\
$P_{2}=$  &$  \sum_{a,b\neq a, c \neq a} (\vec \sigma_a \times \vec \sigma_b)\,\frac{\vec r_{ac} \times \vec r_{ab}}{r_{ac}^3\,r_{ab}^3}  $
 & -0.059\,905  & -0.821\,17 \\
$P_{3}=$ &$  \sum_{a,b\neq a} (\vec \sigma_a \times \vec \sigma_b)\,  i\, p_a^2\,\frac{\vec r_{ab}}{r_{ab}^3}\times \vec p_a  $
 &  0.102\,287  &  1.841\,14  \\
$Q_{9}= $ &$  \sum_{a,b\neq a}  i\, \vec\sigma_a\,p_a^2\, \frac{1}{r_{ab}} \,\vec p_a \times \vec p_b $
 & -0.126\,256  & -3.131\,89 \\
$Q_{10}=$ &$  \sum_{a,b\neq a}  i\, \vec\sigma_a\,p_a^2\, \frac{\vec r_{ab}}{r_{ab}^3}\times (\vec r_{ab} \cdot \vec p_b) \,\vec p_a  $
 & -0.396\,739  & -8.005\,14  \\
$Q_{11}=$ &$  \sum_{a,b \neq a, c \neq b}\vec\sigma_a\,\frac{1}{r_{bc}} \frac{\vec r_{ab}}{r^3_{ab}} \times \vec p_c $
 & -0.114\,547  & -1.441\,91 \\
$Q_{12}=$ &$  \sum_{a,b \neq a, c \neq b}\vec\sigma_a\,\frac{\vec r_{ab} \times \vec r_{bc}}{r_{ab}^3 r_{bc}^3} (\vec r_{bc} \cdot \vec p_c) $
 &  0.053\,650  &  0.682\,88 \\
$P_{4}=$  &$  \sum_{a,b\neq a,c\neq a}(\vec\sigma_a\times\vec\sigma_b)\,\frac{\vec r_{ac}\times\vec r_{bc}}{r_{ac}^3\,r_{bc}^3} $ 
 & 0.059\,905   &  0.821\,17 \\
$Q_{14}=$ &$  \sum_{a,b\neq a}\vec\sigma_a\,\frac{1}{r_{ab}} \,\frac{\vec r_{b}}{r_{b}^3} \times \vec p_a $
 &-0.041\,132   &  0.033\,59 \\
$Q_{15}=$ &$  \sum_{a,b\neq a}\vec\sigma_a\,\frac{\vec r_{b} \times \vec r_{ab}}{r_{b}^3 \, r_{ab}^3}\,(\vec r_{ab} \cdot \vec p_a) $
 &-0.144\,617   & -1.287\,70 \\
$Q_{17}=$ &$  \sum_{a,b\neq a, c \neq b}\vec\sigma_a\,\frac{1}{r_{ab}} \,\frac{\vec r_{bc}}{r_{bc}^3} \times \vec p_a $
 & 0.171\,163   &  2.362\,41 \\
$Q_{18}=$ &$  \sum_{a,b\neq a, c \neq b}\vec\sigma_a\,\frac{\vec r_{ab} \times \vec r_{bc}}{r_{ab}^3 \, r_{bc}^3}\,(\vec r_{ab} \cdot \vec p_a) $
 & 0.065\,529   &  0.700\,97 \\
$Q_{19}=$ &$  \sum_{a,b\neq a}  i\,\vec\sigma_a\,p_b^2\, \frac{1}{r_{ab}} \,\vec p_a \times \vec p_b $
 &-0.224\,280   & -3.050\,65 \\
$Q_{20}=$ &$  \sum_{a,b\neq a}  i\,\vec\sigma_a\,p_b^2\, \frac{\vec r_{ab}}{r_{ab}^3} \times (\vec r_{ab} \cdot \vec p_a) \,\vec p_b $
 &-0.506\,006   & -8.526\,97 \\
\hline
\hline
\end{tabular}
\end{table*}

The second-order contribution is split into parts coming from intermediate
states with specified angular momentum and spin,
\begin{eqnarray}
\bigg \langle H^{(4)}\,\frac{1}{(E-H)'}\,H^{(4)} \bigg \rangle &=&   \label{30} \\
&& \hspace{-4cm} \bigg \langle H^{(4)} \frac{1_{^{2,4}S_o} + 1_{^{2,4}P} + 1_{^{2,4}D_o} + 1_{^4F}}{(E-H)'}\,H^{(4)} \bigg \rangle =\nonumber \\ 
 && \hspace{-4cm}   X_{^2S_o} + X_{^4S_o} + X_{^2P} + X_{^4P} + X_{^2D_o} + X_{^4D_o} + X_{^4F} \nonumber
\end{eqnarray}
where $1_{^{2,4}X}$ is a projection into doublet or quartet state $X$, respectively. 
These contributions are also defined in Table~\ref{tabel2}. 
\begin{table*}[thb]
\renewcommand{\arraystretch}{0.95}
\caption{Second-order contributions to Li and Be$^+$ fine splitting $X = (K_{3/2}-K_{1/2})\, V$, 
the additional prefactor $\{K_{1/2}\,K_{3/2}\}$ is for $J=1/2, 3/2$, correspondingly.
The numerical uncertainties are about $10^{-4}$}
\label{tabel2}
\begin{tabular}{lrr@{\hspace{1.5cm}}w{2.11}w{2.11}}
\hline
\hline
\multicolumn{2}{c}{Contribution} &  $\{K_{1/2},K_{3/2}\}$ & \cent{V_{\rm Li}} & \cent{V_{\rm Be^+}} \\
\hline
$X_{^2S_o}=$ & $\langle\Phi|H^{(4)}_B\,\frac{1_{^2S_o}}{E-H}\,H^{(4)}_B|\Phi\rangle$ &$\{1,0\}$              
& -0.293\,49  &  -1.051\,4(2) \\
$X_{^4S_o}=$ & $\langle\Phi|H^{(4)}_B\,\frac{1_{^4S_o}}{E-H}\,H^{(4)}_B|\Phi\rangle$ &$\{0,2/3\}$            
& -0.443\,91(3) & -1.625\,3  \\[2ex]  
$X_{^2P}=$  &  $\langle\Phi|H^{(4)}_B\,\frac{1_{^2P}}{(E-H)'}\,(2 H^{(4)}_A + H^{(4)}_B)|\Phi\rangle$ &      
&  & \\
          &  $\langle\Phi|H^{(4)}_B\,\frac{1_{^2P}}{(E-H)'}\,[H^{(4)}_A]_r|\Phi\rangle$ &$\{1,-1/2\}$        
& -0.0217(6)    &  -1.435(8) \\
          &  $\delta X_{^2P}$                                                        &$\{1,-1/2\}$           
& -0.086\,80    &  -2.169\,0\\
          & $\langle\Phi|H^{(4)}_B\,\frac{1_{^2P}}{(E-H)'}\,H^{(4)}_B|\Phi\rangle$ &$\{1,1/4 \}$             
& -0.719\,6(6)  &  -5.803(4) \\[2ex]
$X_{^4P}=$  & $\langle\Phi|(H^{(4)}_B + H^{(4)}_C) \,\frac{1_{^4P}}{E-H}\, (H^{(4)}_B + H^{(4)}_C) |\Phi\rangle$ &$  $  & \\
          & $\langle\Phi|H^{(4)}_B\,\frac{1_{^4P}}{E-H}\,H^{(4)}_B |\Phi\rangle$ &$\{1/3,5/6\}  $            
& -0.901\,2(4)  &  -3.475\,5(6)  \\
          & $\langle\Phi|H^{(4)}_C\,\frac{1_{^4P}}{E-H}\,H^{(4)}_C |\Phi\rangle$ &$\{3,3/10\}  $             
& -0.002\,31    &  -0.033\,2  \\
          & $\langle\Phi|H^{(4)}_B\,\frac{1_{^4P}}{E-H}\,H^{(4)}_C|\Phi\rangle$ &$ \{-1,1/2\}  $             
& 0.006\,97     &  0.102\,5  \\[2ex]
$X_{^2D_o}=$ & $\langle\Phi| H^{(4)}_B \,\frac{1_{^2D_o}}{E-H}\,H^{(4)}_B |\Phi\rangle$ &$\{0,3/2\}  $       
& -0.500\,75    &  -1.885\,6(4) \\ [2ex]    
$X_{^4D_o}=$ & $\langle\Phi| (H^{(4)}_B + H^{(4)}_C)\,\frac{1_{^4D_o}}{E-H}\,(H^{(4)}_B + H^{(4)}_C)|\Phi\rangle$ &$  $  \\
          & $\langle\Phi| H^{(4)}_B\,\frac{1_{^4D_o}}{E-H}\,H^{(4)}_B|\Phi\rangle$ &$\{2,1\}  $              
& -0.733\,27(2) &  -2.625\,6(2)\\
          & $\langle\Phi| H^{(4)}_C\,\frac{1_{^4D_o}}{E-H}\,H^{(4)}_C|\Phi\rangle$ &$\{2,1\}  $              
& 0.000\,08     &   0.000\,9\\
          & $\langle\Phi|H^{(4)}_B\,\frac{1_{^4D_o}}{E-H}\,H^{(4)}_C|\Phi\rangle$ &$\{2,-1\}  $              
& 0.000\,00     &  0.000\,1  \\[2ex]
$X_{^4F}=$  &  $\langle\Phi|H^{(4)}_C\,\frac{1_{^4F}}{E-H}\,H^{(4)}_C|\Phi\rangle$ &$\{0,3\}$         	     
&-0.000\,71     &  -0.009\,6 \\[2ex]
$Y_1 =$     &   $\langle\Phi| H^{(4)}_{B} \frac{1}{(E-H)'} \sum_a \delta^3(r_a)|\Phi\rangle$ &               &  & \\
          &   $\langle\Phi| H^{(4)}_{B} \frac{1}{(E-H)'} \sum_a [\delta^3(r_a)]_r|\Phi\rangle$ &$\{1,-1/2\}$ 
& -0.028\,95    &   -0.647\,1 \\
          &   $\delta Y_1$                                                                      &$\{1,-1/2\}$
&  0.007\,99    &   0.186\,0 \\
$Y_2 =$     &   $\langle\Phi| H^{(4)}_{B} \frac{1}{(E-H)'} \sum_{b<a} \delta^3(r_{ab})|\Phi\rangle$ &        &  &  \\
          &   $\langle\Phi| H^{(4)}_{B} \frac{1}{(E-H)'} \sum_{b<a} [\delta^3(r_{ab})]_r|\Phi\rangle$ &$\{1,-1/2\}$
& 0.001\,07     &   -0.002\,5   \\
          &   $\delta Y_2$                                                                      &$\{1,-1/2\}$
& -0.003\,66    &   0.053\,5 \\
\hline
\hline
\end{tabular}
\end{table*}
Most of them can be calculated as they stand. Only the nonsymmetric 
$\langle H_B^{(4)}/(E-H)'\, H_A^{(4)}\rangle$ matrix element needs numerical 
regularization due to the high singularity of $H_A^{(4)}$.
This is done as follows: $H_A^{(4)}$ is transformed to the
regular form by the following transformations
\begin{eqnarray}
4 \pi\, \delta^3(r_a) &=& 4 \pi \, [\delta^3(r_a)]_r - \bigg\{\frac{2}{r_a} ,E-H \bigg\}  \\
4 \pi \, [\delta^3(r_a)]_r &=& \frac{4}{r_a} (E-V) - 2 \sum_b \vec p_b\, \frac{1}{r_a}\, \vec p_b 
\end{eqnarray}
\begin{eqnarray}
4 \pi\, \delta^3(r_{ab}) &=& 4 \pi \, [\delta^3(r_{ab})]_r - \bigg\{\frac{1}{r_{ab}} ,E-H \bigg\}  \\
4 \pi \, [\delta^3(r_{ab})]_r &=& \frac{2}{r_{ab}} (E-V) - \sum_c \vec p_c \,\frac{1}{r_{ab}}\, \vec p_c 
\end{eqnarray}
\begin{eqnarray}
\sum_a p_a^4  &=& \sum_a [p_a^4]_r + 4\,\Big\{V, E-H \Big\} \\
\sum_a [p_a^4]_r &=& 4 \,(E-V)^2 - 2 \sum_{a<b} \vec p_a^{\,2} \, \vec p_b^{\,2}
\end{eqnarray}
The overall regularized form of $H_A^{(4)}$ is
\begin{equation}
H^{(4)}_A = [H^{(4)}_A]_r + \big\{Q_A ,E-H \big\}\,, 
\end{equation}
where
\begin{equation}
Q_A = \frac{Z}{4}\,\sum_a \frac{1}{r_a} - \frac{1}{2}\,\sum_{a<b} \frac{1}{r_{ab}}. \label{38}
\end{equation}
The expectation value of the regularized operator is the same as that without
regularization. What has changed is the second-order matrix element
\begin{eqnarray}
\bigg \langle H_B^{(4)}\,\frac{1}{(E-H)'}\,H_A^{(4)} \bigg \rangle &=& \nonumber \\
&& \hspace{-14ex} \bigg \langle H_B^{(4)}\,\frac{1}{(E-H)'}\,[H_A^{(4)}]_r \bigg \rangle 
+ \frac{\delta X_{^2P}}{2} \label{39}
\end{eqnarray}
where
\begin{eqnarray}
\delta X_{^2P} &=&  2 \,\Big(\langle H_B^{(4)} Q_A \rangle - \langle H_B^{(4)} \rangle \langle Q_A \rangle \Big) \label{40}\\
 &=& \frac{Z}{8} \Bigl(Z \,\langle Q_{21}\rangle   + 2\,\langle Q_{23}\rangle  - \langle Q_{22}\rangle \Bigr) \nonumber \\
 && - \frac{1}{4} \Bigl(Z \,\langle Q_{24}\rangle   + 2\,\langle Q_{26}\rangle  - \langle Q_{25}\rangle \Bigr) \nonumber \\
 && - \bigg(\frac{Z}{4}\, \langle Q_{29}\rangle + \frac{1}{2}\, \langle Q_{31}\rangle - \frac{1}{4}\, \langle Q_{30}\rangle\bigg)
 \nonumber\\&& \times\bigg(\frac{Z}{2} \langle Q_{27}\rangle  - \langle Q_{28}\rangle\bigg)\nonumber
\end{eqnarray}
These additional $Q_i$ operators together with their expectation value are presented in Table \ref{tabel3}. 
\begin{table*}[htb]
\renewcommand{\arraystretch}{0.95}
\caption{Expectation values of additional operators arising from reduction of
the second-order matrix elements, $\langle Q \rangle = K_J \, V$ with
the additional prefactor $K_J=\{1,-1/2\}$ for $J=1/2, 3/2$, correspondingly. All digits are significant.}
\label{tabel3}
\begin{tabular}{rl@{\hspace{1.1cm}}w{5.8}w{5.8}}
\hline
\hline
\multicolumn{2}{c}{Operator} & \cent{V_{\rm Li}} & \cent{V_{\rm Be^+}} \\
\hline
$Q_{21}=$ &$  \sum_{a,c} \vec\sigma_a\,\frac{1}{r_c} \frac{\vec r_a}{r_a^3} \times \vec p_a$                     
& -0.849\,430 &  -9.552\,24 \\
$Q_{22}=$ &$  \sum_{a,b \neq a,c} \vec\sigma_a\,\frac{1}{r_c} \frac{\vec r_{ab}}{r_{ab}^3} \times \vec p_a$      
& -1.432\,170 & -15.223\,86 \\ 
$Q_{23}=$ &$  \sum_{a,b \neq a,c} \vec\sigma_a\,\frac{1}{r_c} \frac{\vec r_{ab}}{r_{ab}^3} \times \vec p_b$      
& 0.242\,656 &   3.250\,48 \\
$Q_{24}=$ &$  \sum_{a,c<d} \vec\sigma_a\,\frac{1}{r_{cd}} \frac{\vec r_a}{r_a^3} \times \vec p_a$                
&-0.400\,085  &  -4.721\,57 \\
$Q_{25}=$ &$  \sum_{a,b \neq a,c<d} \vec\sigma_a\,\frac{1}{r_{cd}} \frac{\vec r_{ab}}{r_{ab}^3} \times \vec p_a$ 
&-0.766\,998  &  -8.959\,65 \\
$Q_{26}=$ &$  \sum_{a,b \neq a,c<d} \vec\sigma_a\,\frac{1}{r_{cd}} \frac{\vec r_{ab}}{r_{ab}^3} \times \vec p_b$ 
& 0.159\,671 &   2.209\,81 \\
$Q_{27} =$ &$  \sum_{a} \frac{1}{r_a}$                                                                           
& 5.638\,906 &   7.898\,02 \\
$Q_{28} =$ &$  \sum_{a<b} \frac{1}{r_{ab}}$                                                                      
& 2.096\,405 &   3.233\,41 \\
$Q_{29}=$ &$  \sum_a\vec\sigma_a\,\frac{\vec r_a}{r_a^3} \times \vec p_a$                                        
& -0.125\,946 &  -0.969\,13 \\
$Q_{30}=$ &$  \sum_{a,b \neq a} \vec\sigma_a\,\frac{\vec r_{ab}}{r_{ab}^3} \times \vec p_a$                      
&-0.224\,641  &  -1.659\,49 \\
$Q_{31}=$ &$  \sum_{a,b \neq a} \vec\sigma_a\,\frac{\vec r_{ab}}{r_{ab}^3} \times \vec p_b$                      
& 0.038\,474 &   0.360\,85 \\
\hline
\hline
\end{tabular}
\end{table*}

The last considered term, the $m\,\alpha^7\,\ln\alpha$ correction from Eq.~(\ref{21}), is represented as
\begin{equation}
E^{(7)}_{\rm log} =  \ln[(Z\,\alpha)^{-2}] \,
\left[\frac{Z}{3}\,\langle D_1 \rangle -\frac{3}{4}\,\langle D_2 \rangle + 2\bigg(\frac{4 Z}{3}\,Y_1 
- \frac{7}{3}\,Y_2\bigg)\right]
\end{equation}
where $D_i$ are defined in Table \ref{tabel1} and $Y_i$ in Table \ref{tabel2}. 
The second-order matrix element $Y$ requires numerical regularization, 
similarly to the one in Eq. (\ref{39}), and is transformed into the following form
\begin{eqnarray}
Y_1 &=& \bigg \langle H_B^{(4)}\,\frac{1}{(E-H)'}\,\sum_a[\delta^3(r_a)]_r \bigg \rangle + \delta Y_1 \\
Y_2 &=& \bigg \langle H_B^{(4)}\,\frac{1}{(E-H)'}\,\sum_{b<a}[\delta^3(r_{ab})]_r \bigg \rangle + \delta Y_2 
\end{eqnarray}
where
\begin{eqnarray}
\delta Y_1 &=& - \frac{1}{2\,\pi}\sum_a\left(
\Bigl\langle \frac{1}{r_a}\,H_B^{(4)}\Bigr\rangle - 
\langle H_B^{(4)}\rangle\, \Bigl\langle \frac{1}{r_a} \Bigr\rangle
\right) \nonumber \\
&=& \frac{1}{8\,\pi}\,\Bigl( -Z\,\langle Q_{21}\rangle + \langle Q_{22}\rangle - 2\,\langle Q_{23}\rangle \nonumber \\ 
&& + \langle Q_{27}\rangle\,(Z\,\langle Q_{29} \rangle -\langle Q_{30}\rangle + 2\,\langle Q_{31}\rangle)\Bigr) \\
\delta Y_2 &=& - \frac{1}{\pi}\sum_{b<a}\left(
\Bigl\langle \frac{1}{r_{ab}}\,H_B^{(4)}\Bigr\rangle - 
\langle H_B^{(4)}\rangle\, \Bigl\langle \frac{1}{r_{ab}} \Bigr\rangle
\right) \nonumber \\
&=& \frac{1}{4\,\pi}\,\Bigl( -Z\,\langle Q_{24}\rangle + \langle Q_{25}\rangle - 2\,\langle Q_{26}\rangle \nonumber \\ 
&& + \langle Q_{28}\rangle\,(Z\,\langle Q_{29} \rangle -\langle Q_{30}\rangle + 2\,\langle Q_{31}\rangle)\Bigr) 
\end{eqnarray}
are expressed in terms of $Q_i$ from Table \ref{tabel3}.

\section{Spin reduction of matrix elements}

The wave function $\Phi^i$ of the $^2$P state in a three-electron system is of
the form 
\begin{equation}
\Phi^i = \frac{1}{\sqrt{6}}\,{\cal A}\big[\phi^i(\vec r_1,\vec r_2,\vec r_3)\,
[\alpha(1)\,\beta(2)-\beta(1)\,\alpha(2)]\,\alpha(3)\big]\,,\label{46}
\end{equation}
where $\cal A$ denotes antisymmetrization and 
$\phi^i(\vec r_1,\vec r_2,\vec r_3)$ is a spatial function with Cartesian
index $i$ that comes from any of the electron coordinates. The normalization we assume is
\begin{equation}
\sum_i \langle\Phi'^i|\Phi^i \rangle = \sum_i
\bigl\langle \phi'^{\,i}(r_1,\,r_2,\,r_3)|{\cal P}
             [c_{123}\,\phi^i(r_1,r_2,r_3)] \bigr\rangle = 1 \label{47}
\end{equation}
where $\cal P$ denotes the sum of all permutations of 1,2, and 3.
The $^2$P$_{1/2}$ and $^2$P$_{3/2}$ wave functions are constructed using
Clebsch-Gordon coefficients. Expectation values with these wave functions
can be reduced to spatial expectation values with algebraic prefactor $K_J$ 
for $J=1/2,3/2$. Namely, the first-order matrix elements with auxiliary 
notation $\{K_{1/2},K_{3/2}\}$ take the form 
\begin{equation}
\langle\Phi'|O|\Phi \rangle =  \{1,1\} \, \bigl\langle \phi'^{\,i}(r_1,\,r_2,\,r_3)|   
Q\, {\cal P} [c_{123}\,\phi^i(r_1,r_2,r_3)] \bigr\rangle  \\
\end{equation}
\begin{eqnarray}
 \langle \Phi'|\sum_a \vec \sigma_a \cdot \vec Q_a |\Phi \rangle &=&   
\{1,-1/2\} \, i\,\epsilon^{ijk}\,\sum_a  \nonumber \\
&&\hspace{-16ex} \Bigl\langle\phi'^{\,i}(r_1,\,r_2,\,r_3)|
Q_a^j\,{\cal P} \Big[c^{Fa}_{123}\,\phi^k(r_1,r_2,r_3)\Big] \Big \rangle
\end{eqnarray}
\begin{eqnarray}
 \langle \Phi'|\sum_{a \neq b} \vec\sigma_a\times \vec\sigma_b\cdot \vec Q_{ab}|\Phi \rangle &=&
 \{1,-1/2\} \, (-2\,\epsilon^{ijk})\!\!\!  \sum_{ab=12,23,31} \nonumber \\
 && \hspace{-26ex} \bigl\langle \phi'^{\,i}(r_1,\,r_2,\,r_3)|
 \,(Q_{ab}^{j}-Q_{ba}^{j})\,{\cal P} \Big[c^{P}_{123}\,\phi^k(r_1,r_2,r_3)\Big] \Big \rangle 
\end{eqnarray}
where $c_{klm}$ coefficients are defined in Table \ref{table6}. 
\begin{table*}[!hbt]
\renewcommand{\arraystretch}{1.0}
\caption{Symmetrization coefficients in matrix elements}
\label{table6}
\begin{ruledtabular}
\begin{tabular}{lrrrrrr}
$(k,l,m)$  & $c_{klm}$ & $c^A_{klm}$ & $c^{F1}_{klm}$ & $c^{F2}_{klm}$ & $c^{F3}_{klm}$ & $c^{P}_{klm}$  \\
\hline 
 \\
 $(1,2,3)$    &   2   &   1 &    0    &    0    &     2   &     0   \\
 $(1,3,2)$    &  -1   &  -1 &    1    &   -1    &    -1   &     1   \\
 $(2,1,3)$    &   2   &   1 &    0    &    0    &     2   &     0   \\
 $(2,3,1)$    &  -1   &  -1 &   -1    &    1    &    -1   &    -1   \\
 $(3,1,2)$    &  -1   &   1 &    1    &   -1    &    -1   &     1   \\
 $(3,2,1)$    &  -1   &  -1 &   -1    &    1    &    -1   &    -1   \\
  \end{tabular}
\end{ruledtabular}
\end{table*}

The spin reduction of the second-order matrix elements is more complicated.
We shall first introduce the following auxiliary functions,
\begin{eqnarray}
\Psi^i &=& Q\,{\cal P}[c_{123}\,\phi^i(r_1,r_2,r_3)]\\
\Psi^{ij} &=& \sum_a Q_a^i\,{\cal P} \Big[  c^{Fa}_{123}\,\phi^j(r_1,r_2,r_3)\Big]\\
\Psi_A^{ij} &=& {\cal P} \Big[c^{A}_{123}\,(Q_1^i - Q_2^i)\,\phi^j(r_1,r_2,r_3)\Big]\\
\Psi_A^{ijk} &=& {\cal P} \Big[c^{A}_{123}\,(Q_{13}^{ij} - Q_{23}^{ij})\,\phi^k(r_1,r_2,r_3)\Big]
\end{eqnarray}
Then, the spin-reduced second-order matrix elements are
\begin{widetext}
\begin{eqnarray}
\langle\Phi|\sum_a \vec \sigma_a \cdot \vec Q_a \,\frac{1_{^2S_o}}{E-H}\,\sum_b \vec \sigma_b \cdot \vec Q_b |\Phi\rangle &=& 
\frac{\{1,0\}}{6}\,\Bigl\langle \Psi^{ii}\,\frac{1_{^2S_o}}{E-H}\,\Psi^{jj} \Big \rangle
\end{eqnarray}
\begin{eqnarray}
\langle\Phi|\sum_a \vec \sigma_a \cdot \vec Q_a \,\frac{1_{^4S_o}}{E-H}\,
\sum_b \vec \sigma_b \cdot \vec Q_b |\Phi\rangle &=& \frac{\{0,2/3\}}{6}\,
\Bigl\langle\Psi_A^{ii}\frac{1_{^4S_o}}{E-H}\Psi_A^{jj} \Big \rangle
\end{eqnarray}
\begin{eqnarray}
\langle\Phi| Q \,\frac{1_{^2P_o}}{(E-H)'}\,Q |\Phi\rangle &=& \frac{\{1,1\}}{6}\,
\Bigl\langle\Psi^i\,\frac{1_{^2P_o}}{(E-H)'}\, \Psi^{i}\Bigr \rangle
\end{eqnarray}
\begin{eqnarray}
\langle\Phi| Q \,\frac{1_{^2P_o}}{(E-H)'}\,\sum_a\vec \sigma_a \cdot \vec Q_a |\Phi\rangle &=& \frac{\{1,-1/2\}}{6}\,
\Bigl\langle\Psi^i\,\frac{1_{^2P_o}}{(E-H)'}\, i\,\epsilon^{ijk}\,\Psi^{jk}\Bigr \rangle
\end{eqnarray}
\begin{eqnarray}
\langle\Phi|\sum_a \vec \sigma_a \cdot \vec Q_a \,\frac{1_{^2P_o}}{(E-H)'}
\,\sum_b\vec \sigma_b \cdot \vec Q_b |\Phi\rangle &=& \frac{\{1,1/4 \}}{6}\,
\Bigl\langle i\,\epsilon^{ijk}\Psi^{ij}\frac{1_{^2P_o}}{(E-H)'}\,i\,\epsilon^{lmk}\Psi^{lm}\Big \rangle
\end{eqnarray}
\begin{eqnarray}
\langle\Phi|\sum_a \vec \sigma_a \cdot \vec Q_a \,\frac{1_{^4P_o}}{E-H}\,
\sum_b \vec \sigma_b \cdot \vec Q_b |\Phi\rangle &=& \frac{\{1/3,5/6\}}{6}\,
\Bigl\langle i\,\epsilon^{ijk} \Psi_A^{ij}\frac{1_{^4P_o}}{E-H}\,i\,\epsilon^{lmk}\Psi_A^{lm} \Big \rangle
\end{eqnarray}
\begin{eqnarray}
\langle\Phi|\sum_{a<b} \sigma_a^i \sigma_b^j\,Q_{ab}^{ij} \,\frac{1_{^4P_o}}{E-H}\,
\sum_{c<d} \sigma_c^i \sigma_d^j\,Q_{cd}^{ij} |\Phi\rangle &=& \frac{\{3,3/10\}}{6}\,
\Bigl\langle\Psi_A^{ijj}\frac{1_{^4P_o}}{E-H}\Psi_A^{ikk} \Big \rangle
\end{eqnarray}
\begin{eqnarray}
\langle\Phi| \sum_a \vec \sigma_a \cdot \vec Q_a \,\frac{1_{^4P_o}}{E-H}\,
\sum_{b<c} \sigma_b^i \sigma_c^j\,Q_{bc}^{ij} |\Phi\rangle &=& \frac{\{-1,1/2\}}{6}\,
\Bigl\langle i\,\epsilon^{jml}\Psi_A^{ml}\frac{1_{^4P_o}}{E-H}\Psi_A^{jkk} \Big \rangle
\end{eqnarray}
\begin{eqnarray}
\langle\Phi|\sum_a \vec \sigma_a \cdot \vec Q_a \,\frac{1_{^2D_o}}{E-H}\,
\sum_b\vec \sigma_b \cdot \vec Q_b |\Phi\rangle &=& \frac{\{0,3/2\}}{6}\,
\Bigl\langle\Psi^{ji}\frac{1_{^2D_o}}{E-H}\Psi^{ij}\Big \rangle
\end{eqnarray}
\begin{eqnarray}
\langle\Phi|\sum_a \vec \sigma_a \cdot \vec Q_a \,\frac{1_{^4D_o}}{E-H}\,
\sum_b \vec \sigma_b \cdot \vec Q_b |\Phi\rangle &=& \frac{\{2,1\}}{6}\,
\Bigl\langle\Psi_A^{ji}\frac{1_{^4D_o}}{E-H}\Psi_A^{ij} \Big \rangle
\end{eqnarray}
\begin{eqnarray}
\langle\Phi|\sum_{a<b} \sigma_a^i \sigma_b^j\,Q_{ab}^{ij} \,\frac{1_{^4D_o}}{E-H}\,
\sum_{c<d} \sigma_c^i \sigma_d^j\,Q_{cd}^{ij} |\Phi\rangle &=& \frac{\{2,1\}}{6}\,
\Bigl\langle i\,\epsilon^{ijk}\Psi_A^{lij}\frac{1_{^4D_o}}{E-H}\,i\,\epsilon^{mnk}\Psi_A^{lmn} \Big \rangle
\end{eqnarray}
\begin{eqnarray}
\langle\Phi| \sum_a \vec \sigma_a \cdot \vec Q_a \,\frac{1_{^4D_o}}{E-H}\,
\sum_{b<c} \sigma_b^i \sigma_c^j\,Q_{bc}^{ij} |\Phi\rangle &=& \frac{\{2,-1\}}{6}\,
\Bigl\langle \Psi_A^{lk} \frac{1_{^4D_o}}{E-H}\,i\,\epsilon^{kmn}\Psi_A^{lmn}\Big \rangle 
\end{eqnarray}
\begin{eqnarray}
\langle\Phi|\sum_{a<b} \sigma_a^i \sigma_b^j\,Q_{ab}^{ij} \,\frac{1_{^4F_o}}{E-H}\,
\sum_{c<d} \sigma_c^i \sigma_d^j\,Q_{cd}^{ij} |\Phi\rangle &=& \frac{\{0,3\}}{6}\,
\Bigl\langle\Psi_A^{kji}\frac{1_{^4F_o}}{E-H}\Psi_A^{ijk} \Big \rangle
\end{eqnarray}
\end{widetext}
These formulas, including $K_J$ coefficients, have been obtained with a computer symbolic program.

\section{Numerical calculations}
The spatial function $\phi$ in Eq.~\eqref{46} is represented as a linear combination of
the Hylleraas \cite{wang} or the explicitly correlated Gaussians functions \cite{gauss} 
\begin{eqnarray}
\phi &=& \left\{\begin{array}{l}
                          e^{-\alpha_1 r_1^2 -\alpha_2 r_2^2 -\alpha_3 r_3^2 -\alpha_{12} r_{12}^2 -\alpha_{13} r_{13}^2 -\alpha_{23} r_{23}^2} \\
                          e^{-\alpha_1 r_1 -\alpha_2 r_2 -\alpha_3 r_3}\, r_{23}^{n_1}\, r_{31}^{n_2}\, r_{23}^{n_3}\,
                          r_1^{n_4}\,r_2^{n_5}\,r_3^{n_6}
                       \end{array}\right. \label{68}
\end{eqnarray}
In the Hylleraas basis we use six sectors with different values of nonlinear parameters $w_i$
and a maximum value of $\Omega \equiv n_1+n_2+n_3+n_4+n_5 = 12$; details are presented in Refs. \cite{lit_fine_yd, lit_fine}.
In Gaussian basis we use $N=256, 512$, $1024$, and $2048$ functions with well-optimized nonlinear parameters for each basis function separately.
The accuracy achieved for nonrelativistic energies is about $10^{-13}$ 
in Hylleraas and $10^{-11}$ in Gaussian bases.

These nonrelativistic wave functions are used in evaluation of matrix elements. 
Most of the $Q$ and $P$ operators in Tables \ref{tabel1} and \ref{tabel3} are 
intractable with present algorithms with Hylleraas functions due to difficulties with 
integrals with inverse powers of electron distances, 
but also due to very lengthy expressions in terms of Hylleraas integrals. 
Thus, we calculate them using Gaussian functions; however,  with some exceptions.
There are operators $Q_1$, $Q_2$, and $Q_4$, the expectation value 
of which is very slowly convergent. Namely, the accuracy achieved is as low as 
$10^{-2}-10^{-3}$ with as many as 2048 well-optimized Gaussian functions. 
So, to avoid loss of numerical accuracy, we represent these operators 
as the sum of the singular $D$-part and the regular $R$-part. The singular D-part,
numerically dominating, is calculated with Hylleraas functions,
while the regular R-part, free of singularities, is calculated with a Gaussian basis.
This leads to significant improvements in accuracy, so
the numerical uncertainties do not affect theoretical predictions for the fine structure.  
Numerical results for all first-order matrix elements obtained with 
the largest basis are presented in Table \ref{tabel1} and \ref{tabel3}.
The achieved precision is at least $10^{-5}$, which is one digit better
in comparison to second-order matrix elements described in the following.

The evaluation of second-order matrix elements is much more computationally demanding.
First of all, they are obtained only in the Gaussian basis, due to its high flexibility.
The resolvent $1/(E-H)$ for each angular momentum is represented in terms of 
functions with the appropriate Cartesian prefactor, as follows
\begin{equation}
\phi_{S_o} = \epsilon_{ijk} r_a^i r_b^j r_c^k \; \phi  \label{69}
\end{equation}
\begin{equation}
\phi_{P_o}^{i} = r_a^i \; \phi \label{70}
\end{equation}
\begin{equation}
\phi_{D_o}^{ij} = \bigg[ \frac{\epsilon_{ikl}}{2}\,r_c^j  + \frac{\epsilon_{jkl}}{2}\,r_c^i  
-\frac{\delta^{ij}}{3} \epsilon_{mkl}\,r_c^m \bigg] r_a^k\,r_b^l\phi \label{71}
\end{equation}
\begin{eqnarray}
\phi_{F_o}^{ijk} &=& \bigg[ 
   \frac{r_a^i}{6} (r_b^j r_c^k + r_c^j r_b^k) + 
   \frac{r_b^i}{6} (r_a^j r_c^k + r_c^j r_a^k) \nonumber \\ &&+ 
   \frac{r_c^i}{6} (r_a^j r_b^k + r_b^j r_a^k)
\nonumber \\
&& - \frac{\delta^{jk}}{15} \big(r_a^i r_b^l r_c^l + r_b^i r_a^l r_c^l + r_c^i r_a^l r_b^l \bigr)
 \nonumber \\
	 &&   - \frac{\delta^{ki}}{15} \big(r_a^j r_b^l r_c^l + r_b^j r_a^l r_c^l + r_c^j r_a^l r_b^l \bigr) \nonumber \\
         && - \frac{\delta^{ij}}{15} \big(r_a^k r_b^l r_c^l + r_b^k r_a^l r_c^l + r_c^k r_a^l r_b^l \bigr)
               \bigg] \phi\label{72}
\end{eqnarray}
where subscripts $a,b$, and $c$ refer to any of the electrons including the same one.
Nonlinear parameters for intermediate states are extensively optimized for each second-order symmetric 
matrix element. Moreover, one takes all possible representations of angular factors for intermediate states
in appropriate proportions to ensure the completeness of the basis.
Most importantly, the number of Gaussian functions for intermediate states is chosen to be sufficiently high
to saturate the matrix element. Namely, for a given size $N$ of the external wave function, 
we use $3/2\,N$ elements for all $D_o$- and quartet $F_o$-states, $N$ elements for quartet $P_o$-states, and
$1/2\,N$ for $S_o$-states. Among all matrix elements, the most demanding in terms of optimization 
was that with intermediate states of symmetry $^2P_o$, as the external wave function. 
Here, the basis set for the resolvent is divided into two sectors. 
The first sector is built of the known basis functions with the nonlinear
parameters determined in the minimization of $E(2^2P)$.
For this purpose we took one of the previously generated basis sets of $\Psi$
of size equal to $N/2$. The nonlinear parameters of this basis remain fixed during the optimization
in order to ensure the  accurate representation of the states orthogonal to $\Psi$.
The second sector, of size equal to $3/2 N$ or $N$ for the matrix element 
involving $H_B$ or $[H_A]_r$, respectively, 
consists of basis functions that undergo optimization. For the asymmetric matrix elements
the basis is combined from two corresponding symmetric ones. 

The most computationally demanding matrix element was 
the $\langle [H_A]_r\, 1/(E-H)'\,H_B\rangle$ term, 
and it has the slowest numerical convergence in the Gaussian basis.
Numerical results for matrix elements are summarized in Table \ref{tabel2}.
The achieved precision is about $10^{-4}$, one digit less than the first-order matrix elements. 
\begin{table}[htb]
\renewcommand{\arraystretch}{1.0}
\caption{Summary of $m\,\alpha^6$ contributions to fine splitting.}
\label{tabel4}
\begin{tabular}{r@{\hspace{1.1cm}}w{4.10}w{4.10}}
\hline
\hline
Contribution               & \centt{Li} & \centt{Be$^+$} \\
\hline
$X_{^2S_o}$                 &  0.293\,49     &  1.051\,4(3) \\
$X_{^4S_o}$                 & -0.295\,94(2)  & -1.083\,5(1)\\
$X_{^2P_o}$                 &  0.735\,0(18)  & 11.912(24) \\
$X_{^4P_o}$                 & -0.423\,5(2)   & -1.340\,5(3)  \\
$X_{^2D_o}$                 & -0.751\,13(2)  & -2.828\,4(6) \\
$X_{^4D_o}$                 &  0.733\,34(2)  &  2.625\,7(2) \\
$X_{^4F_o}$                 & -0.002\,13     & -0.028\,9(1) \\[1ex]
total second order          &  0.289\,16(19) & 10.308(24)    \\[1ex]
$\delta H_1$      	    & -0.445\,2(16)  & -13.160(3) \\
$\delta H_2$      	    &  0.026\,31     &  0.577\,8 \\
$\delta H_4$      	    &  0.105\,63     &  2.405\,5(1) \\
$\delta H_5$      	    &  0.150\,52     &  3.186\,6 \\
$\delta H_6$      	    & -0.011\,60     & -0.130\,7 \\
$\delta H_7$      	    & -0.027\,83     & -0.757\,8(3) \\[1ex]
total first order 	    & -0.202\,1(16)  & -7.879(3)\\[1ex]
total $m\,\alpha^6$ 	    & 0.087\,1(24)   &  2.429(24) \\
\hline  
\hline
\end{tabular}
\end{table}
In addition, we observe significant cancellations between $S=1/2$ and $S=3/2$ intermediate
states, and between the first- and second-order terms, see Table \ref{tabel4}.
The final numerical result for the $m\,\alpha^6$ contribution
in Table \ref{tabel4} is relatively quite small. 
Regarding the $m\,\alpha^7$ contribution, the second-order term $Y$ is numerically
dominant, and contributions from $D_i$ terms are an order of magnitude smaller.
Altogether this correction is only three times smaller than the $m\,\alpha^6$ contribution,
which is certainly not negligible.

\section{Summary}
We have performed accurate calculations of the fine structure in Li and Be$^+$ using the nonrelativistic QED approach combined with explicitly correlated
basis functions. Relativistic and QED corrections 
are represented in terms of matrix elements of effective operators, 
which are calculated with a highly accurate nonrelativistic wave function. 
Numerical results are summarized in Table \ref{tabel5}. 
\begin{table*}[thb]
\renewcommand{\arraystretch}{1.0}
\caption{Fine splitting of 2P-states in Li and Be$^+$ in units of MHz. 
         $\delta E_{\rm fs}$ is the hyperfine mixing correction. The uncertainty due to neglected terms
         is estimated to be 50\% of $E_{\rm fs log}^{(7,0)}$ }
\label{tabel5}
\begin{ruledtabular}
\begin{tabular}{rw{8.9}cw{8.7}c}
         & \centt{$^7$Li} & \centt{Ref.} & \centt{$^9$Be$^+$} & \centt{Ref.}\\
$E_{\rm fs}^{(4,0)}$            & 10\,053.707(8)     & \cite{lit_fine}   & 197\,039.15(8)    & \cite{lit_fine} \\
$E_{\rm fs}^{(4,1)}$            &      -2.389        & \cite{lit_fine}   &      -21.27       & \cite{lit_fine} \\
$E_{\rm fs}^{(6,0)}$            &       1.63(5)      & \cite{lit_fs}     &       45.4(4)   & 
\\
$E_{\rm fs log}^{(7,0)}$        &       0.15(7)      &                    &        4.6(2.3)   & \\
$\delta E_{\rm fs}$             &       0.159 & \cite{lit_fine}          &        0.03       & \cite{lit_fine}\\
$E_{\rm fs}$(theo)              &   10\,053.25(9)    & \cite{lit_fs}     &  197\,068.0(2.4)  &  \cite{noerters}  \\[1ex]
$E_{\rm fs}$(theo)              & 10\,052.(43)       & Yan {\em et al.} \cite{yan_rel} & 197\,024.(150) & Yan {\em el al.} \cite{yan_rel} \\
$E_{\rm fs}$(exp)               & 10\,053.310(17)    & Brown {\em et al.} \cite{brown} & 197\,063.48(52) &  N{\"o}rtersh{\"a}user {\em et al.}\cite{noerters}\\
$E_{\rm fs}$(exp)               & 10\,053.24(22)     & Brog {\em et al.} \cite{brog}   & 197\,144.         & Ralchenko {\em et al.}\cite{ralchenko} \\
$E_{\rm fs}$(exp)               & 10\,053.184(58)    & Orth {\em et al.} \cite{orth}   & 197\,150.(64)     & Bollinger {\em et al.} \cite{bollinger1985} \\
$E_{\rm fs}$(exp)               & 10\,053.119(58)    & Noble {\em et al.} \cite{noble}\\
\end{tabular}
\end{ruledtabular}
\end{table*}
We observe an agreement with the experimental values. However,
our result for Li lies below, while for Be$^+$ 
above experiments of \cite{brown} and \cite{noerters} respectively.
As the sign of all corrections is the same for Li and Be$^+$, 
this may suggest that one of these experiments underestimated
its uncertainty.  

The extension of presented computational approach to other systems with more electrons is problematic,
due to a lack of formulas for the four-electron Hylleraas integrals.  
Therefore, achieving similar accuracy for the four electron systems
would be very challenging.

\section*{Acknowledgments}
The authors acknowledge the support of NCN grant 2012/04/A/ST2/00105, and by PL-Grid Infrastructure.

\appendix

\section{Quantum mechanics of three identical particles} 
Consider a wave function of three identical particles
$\phi(\vec r_1,\vec r_2,\vec r_3)$. Eigenstates of the nonrelativistic Hamiltonian
can be classified by representation of the permutation group $S_3$.
Two of them, {\em id} and {\em sgn}, are one dimensional, and 
the third is two dimensional. 
The wave functions corresponding to one-dimensional representations are
\begin{eqnarray}
\psi_S(\vec r_1,\vec r_2,\vec r_3) &=& \frac{1}{\sqrt{6}}\,[
\phi(\vec r_1,\vec r_2,\vec r_3) + \phi(\vec r_2,\vec r_3,\vec r_1)  + \\ 
&& \hspace{-2.5cm} \phi(\vec r_3,\vec r_1,\vec r_2) +\phi(\vec r_2,\vec r_1,\vec r_3) + \phi(\vec r_3,\vec r_2,\vec r_1) + \phi(\vec r_1,\vec r_3,\vec r_2)]\nonumber
\end{eqnarray}
and
\begin{eqnarray}
\psi_A(\vec r_1,\vec r_2,\vec r_3) &=& \frac{1}{\sqrt{6}}\,[
\phi(\vec r_1,\vec r_2,\vec r_3) + \phi(\vec r_2,\vec r_3,\vec r_1) + \\ 
&&  \hspace{-2.5cm} \phi(\vec r_3,\vec r_1,\vec r_2)-\phi(\vec r_2,\vec r_1,\vec r_3) - \phi(\vec r_3,\vec r_2,\vec r_1) - \phi(\vec r_1,\vec r_3,\vec r_2)] \nonumber
\end{eqnarray}
In order to construct the wave functions corresponding to the two-dimensional
representation, let us consider the spin-dependent wave function 
for a three-electron system for the total spin $S=1/2$
\begin{eqnarray}
\Phi &=& \frac{1}{\sqrt{6}}\,{\cal A}\big[\phi(\vec r_1,\vec r_2,\vec r_3)\,
[\alpha(1)\,\beta(2)-\beta(1)\,\alpha(2)]\,\alpha(3)\big] \nonumber \\&=&
 \frac{1}{\sqrt{6}}[
\alpha(1)\,\beta(2)\,\alpha(3)\,\psi_1 +
\beta(1)\,\alpha(2)\,\alpha(3)\,\psi_2 + \nonumber \\
&& \alpha(1)\,\alpha(2)\,\beta(3)\,\psi_3]
\end{eqnarray}
where $\cal A$ denotes antisymmetrization, and
\begin{eqnarray}
\psi_1 &=&\phi(\vec r_1,\vec r_2,\vec r_3) - \phi(\vec r_2,\vec r_3,\vec r_1)
+  \phi(\vec r_2,\vec r_1,\vec r_3) - \nonumber \\
&&\phi(\vec r_3,\vec r_2,\vec r_1) \\
\psi_2 &=&\phi(\vec r_3,\vec r_1,\vec r_2) - \phi(\vec r_2,\vec r_1,\vec r_3)
- \phi(\vec r_1,\vec r_2,\vec r_3) + \nonumber \\
&& \phi(\vec r_1,\vec r_3,\vec r_2)\\
\psi_3 &=&\phi(\vec r_2,\vec r_3,\vec r_1) - \phi(\vec r_3,\vec r_1,\vec r_2)
+ \phi(\vec r_3,\vec r_2,\vec r_1) - \nonumber \\
&& \phi(\vec r_1,\vec r_3,\vec r_2)
\end{eqnarray}
$\psi_i$ functions form a two-dimensional representation of $S_3$, $\sum_i \psi_i = 0$. 

Let us denote the standard matrix element
\begin{equation}
\langle\phi'|\phi\rangle_S = \bigl\langle \phi'^{\,i}(r_1,\,r_2,\,r_3)|{\cal P} [c_{123}\,\phi^i(r_1,r_2,r_3)] \bigr\rangle
\end{equation}
where $\cal P$ denotes the sum of all permutations of 1,2, and 3. Then
\begin{equation}
\langle\Phi'|\Phi\rangle =  \langle\phi'|\phi\rangle_S
\end{equation}
and the scalar products between $\psi_i$ is
\begin{equation}
\langle\psi'_i|\psi_j\rangle =  \langle\phi'|\phi\rangle_S\,(-1+3\,\delta_{ij})
\end{equation}
The two orthogonal and normalized functions can be chosen as
$\psi_I = \psi_1/\sqrt{2}$ and $\psi_{II} = (\psi_2-\psi_3)/\sqrt{6}$.

The first-order matrix elements of the spin-independent operator $Q$ are
\begin{equation}
\langle\Phi|Q|\Phi\rangle = \frac{1}{6}\langle\psi_i|Q|\psi_i\rangle = \langle\phi|Q|\phi\rangle_S,
\end{equation}
and the second-order matrix elements with $Q_1$ and $Q_2$ are
\begin{eqnarray}
\langle\Phi|Q_1\,\frac{1}{E-H}\,Q_2|\Phi\rangle &=& \frac{1}{6}\langle\psi_i|Q_1\,\frac{1}{E-H}\,Q_2|\psi_i\rangle
\nonumber  \\ &=&
\langle\phi|Q_1\frac{1}{E-H}\,Q_2|\phi\rangle_S
\end{eqnarray}
In the numerical evaluation of second-order matrix elements with doublet $S=1/2$ intermediate states, 
the resolvent $1/(E-H)$ is represented on the basis of functions of proper $S_3$ symmetry, namely
$\psi_I$ and $\psi_{II}$
\begin{eqnarray}
\langle\psi_I^k|E-H|\psi_I^l\rangle &=& \langle\psi_{II}^k|E-H|\psi_{II}^l\rangle \\
&=& \langle\phi^k|E-H|\phi^l\rangle_S  = E\,{\cal N}_{kl} - {\cal H}_{kl}\nonumber
\end{eqnarray}
Hence, the second-order matrix element using Eq. (A7) becomes
\begin{widetext}
\begin{eqnarray}
\langle\Phi|Q_1\,\frac{1}{E-H}\,Q_2|\Phi\rangle &=&
\frac{1}{6}\langle\psi_i|Q_1|\psi_I^k\rangle\,(E\,{\cal N}-{\cal H})^{-1}_{kl}\,\langle\psi_I^l|Q_2|\psi_i\rangle  +
\frac{1}{6}\langle\psi_i|Q_1|\psi_{II}^k\rangle\,(E\,{\cal N}-{\cal H})^{-1}_{kl}\,\langle\psi_{II}^l|Q_2|\psi_i\rangle
\\ &=&
\langle\phi|Q_1|\phi^k\rangle_S\,(E\,{\cal N}-{\cal H})^{-1}_{kl}\,\langle\phi^l|Q_2|\phi\rangle_S 
\nonumber \\ 
&=& \langle{\cal P} [c_{123}\,\phi(r_1,r_2,r_3)|Q_1|\phi^k\rangle\,
(E\,{\cal N}-{\cal H})^{-1}_{kl}\,\langle\phi^l|Q_2|{\cal P} [c_{123}\,\phi(r_1,r_2,r_3)\rangle
\nonumber
\end{eqnarray}
\end{widetext}
and the last form is used in the numerical calculations.
\end{document}